# Radio Frequency Speed Bumps for Near-Zero Speed Zone Control in Nigeria

Olaide Ayodeji Agbolade
The Department of Electrical and Electronics Engineering, Federal University of Technology, Akure, Ondo State, Nigeria
(oaagbolade@futa.edu.ng)

*Abstract*-Over speeding is an important problem in Nigeria. The Nigerian Federal Road Safety Commission report shows that over speeding is directly responsible for over 50 percent of road crashes in the country. Speed bumps particularly on highways have not proved to be the solution; they rather contribute more to the problem. Based the advent of the dedicated short range communications (DSRC) systems defined by the IEEE 802.11p standard, we proposed a radio frequency based speed bump to mimic the operation of traditional speed breakers. The model consists of a roadside unit and an in-vehicle unit that works with the already installed speed limiter in vehicles. In our work, we showed that the RF based model is able to decelerate the vehicle at a recommended rate of 10 m/s$^2$ over a convenient stopping distance of 80 m. Our work also showed that signal strength based adaptive speed limiter systems can be used as a reliable substitute to GPS thus eradicating the need for additional hardware in the design.

*Keywords-* Speed Bump, Intelligent Speed Adaptation System, Connected Autonomous Vehicle

## I. INTRODUCTION

Over speeding is a significant problem particularly in developing countries with inadequate infrastructure to track and adequately control vehicular traffic. These menace according to a Federal Road Safety Commission (FRSC) Nigeria 2015 report (as cited in [1]) is directly responsible for over 50 percent of all road crashes in Nigeria. A world health organization (WHO) 2017 road traffic accidents deaths report on Nigeria [2] stated that the country's death rate is 24.75 per 100,000 population thus making Nigeria the 58th in the world accident fatality list. Road accident kills at least 2 people every four hours in Nigeria [3]. In Africa as a whole, accident related death rate is put 26.6 deaths per 100,000 people with Zimbabwe being worse off with 74.5 deaths for every 100,000 inhabitants [3] which, by every standard is far too high considering the fact that the global average is 17.4.

One common way of checking over speeding is through the use of speed breakers or speed bumps. These bumps are placed at strategic places where drivers' speed are to be kept minimal. Unfortunately, speed breakers present new set of challenges for road users. The situation for developing countries like Nigeria is even worse because of proliferation of indiscriminate speed breakers that are unprofessionally built. Worse still, most of the speed bumps are put on roads without associated road safety signs to warn drivers. At night or during periods with low visibility, this can be extremely dangerous particularly for drivers who are unfamiliar with the road. A study conducted by [4],[5] revealed that speed bumps on highways in fact do not stop accidents, rather, they cause them. The study further revealed that bumps cause enormous damages to vehicles. It is not uncommon to find bolts, nuts, screws and other objects falling off vehicles at locations where road bumps are installed. A separate study conducted by [6] also identified wear and tear of vehicles, tyre puncturing, increase stress, body aches, delays and environmental pollution as some of the effect of bumps on road users in Nigeria. Correlation between travelling on bumpy roads and miscarriage in women in the first trimester of pregnancy was also reported by [5],[7]. Also, most modern sport cars have very little ground clearance. Road bumps presents a more serious threat to these types of vehicles who suffer more recurrent damages to their silencers and underparts due to frequent travelling on bumps. From the foregoing, it is agreeable that traditional speed breakers have very minimal advantage. Furthermore, the need for speed breakers may also be time based. They may be needed at certain hours of the day and unnecessary at other times. For instance, a particular road may require bumps during the day when school children are likely to use the road more often. At night, this may not be necessary.

With the advent of IoT and dedicated short-range communications (DSRC) systems as defined by the IEEE 802.11p standard, a cost effective approach can be taken to ensure road safety and comfort for all road users without unnecessarily bumping them. This has led to the development of adaptive speed breaking. One of such systems is the Actibump system which has been successfully implemented in Sweden [8]. The design of the system includes the integration of smart powered equipment that is capable of creating bumps on the road by lowering that segment of the road by a few centimeters when its radar system senses the approach of a vehicle travelling above the recommended speed limit. While vehicles travelling above speed limit experiences this bump that forces it to lower its speed as it travels across the Actibump system, a vehicle operating below the speed limit is guaranteed a smooth surface to travel on. SmartBump [9] is another similar system integrated into roads to mechanically form bumps when it detects over speeding in approaching vehicle. While these systems appears promising, it is difficult to imagine them been appealing enough to be adopted in Africa and Nigeria in particular. First, substantial amount of energy is required to power the mechanical systems used in these designs. Secondly, substantial road construction work is needed for their installation. The suitability of these systems for highways where vehicles may be travelling at speed above 100 km/h is also in doubt.

In 2016, the FRSC in its first attempt at utilizing technology to address the menace of over speeding in Nigeria mandated all trucks and buses to install speed limiting devices on their vehicles [10]. The directive was expected to be extended to all other vehicle categories plying the nation's road. Speed limiting devices operates by automatically cutting off fuel and air supply to vehicle's engine when it reaches a pre-set speed limit. From February 2017, the law has been enforced and no vehicle is expected to ply any road in Nigeria without having a functional speed limiter installed. While there are several accompanying technical snags with the implementation of the speed limiting devices in Nigeria, there also exist a huge potential of eliminating the need for speed bumps on the nation's road if integrated with the appropriate technology. This is the primary objective of this work. An illustration of this work is shown in Fig. 1. To achieve this, an IoT and RF based speed adaptation system design is proposed. The proposed model is designed to be suitable particularly for dual carriageways. A simplex communication mode has been employed to reduce complexity and cost of implementation. The use of GPS for ranging together with its associated cost and shortcomings have also been avoided in the system design. The system operates through medium range low power radio frequency transmitter which sends out radio waves that is picked up by a receiver (in vehicle unit) inside the vehicle when within range. The moment a successful communication is established, first, the intelligent speed adaption system in the car determines if the car is moving towards or moving away from the electronic speed breaking system. Next, using received signal strength based distance ranging estimates, the vehicle initiates a systemic breaking at a rate of $10\ m/s^2$. While there are several variabilities that can be factored in to this work, the overarching goal is to electronically reduce the speed of the car to near zero over a convenient stopping distance at the low speed zone, thus eliminating the need for installation of bumps on the road.

## II. BACKGROUND STUDY AND RELATED WORKS

Over speeding related accidents claim thousands of lives annually while many other become permanently incapacitated through various degrees of injuries. Government at all levels have the responsibility of enforcing speed restrictions to limit this threat. However, the thousands of fines still being paid globally for over speeding related offenses indicate that many of such measures offer limited restraints to offenders.

Road bump is one of the most intrusive means of forcing drivers to limit their speed. The dimension of the bump normally determines the maximum convenient speed at which drivers can travel. Most road bumps in Nigeria are either made of asphalt, rubber or concrete and usually span the entire length of the road. Their shapes ranges from circular, parabolic to wave-like shape. The wave-like shaped bumps are usually the most inconvenient to travel on because they are made by combining three of more circular shaped bumps. The width of the bumps ranges from 15 cm for small ones to over 70 cm for large ones with height ranging from about 4 cm to 12 cm. In most countries road bumps are limited to roads in residential areas and urban centres where speed is never expected to exceed 40 miles/hr [11]. However in Nigeria, it is not uncommon to find road bumps even on highways. Due to poor planning, instances of highways passing through rural settlement are very common.

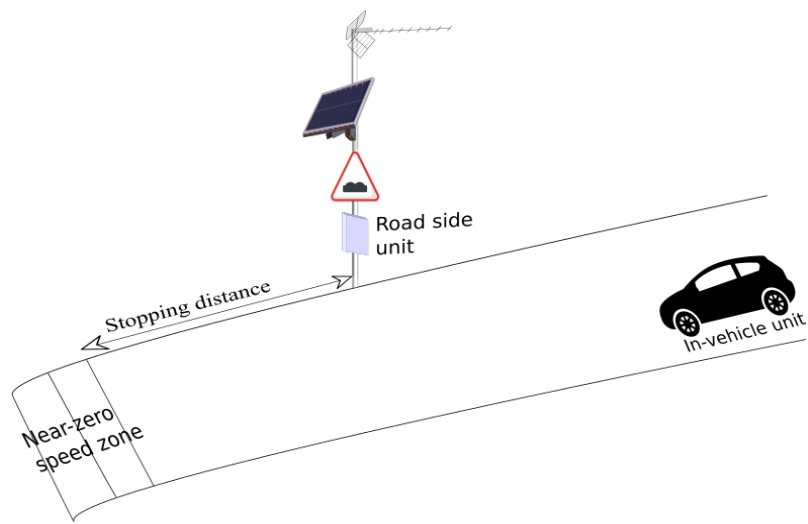

Figure 1. Illustration diagram of proposed system

Most times, in order for the community dwellers to protect themselves from over speeding drivers, road bumps are indiscriminately constructed sometimes with no road sign to warn unsuspecting drivers.

Several research efforts have been put into using technology to limit excessive speed on roads particularly in today's world of connected autonomous vehicles (CAVs) and vehicle to anything systems (V2X) [12]. Over the past couple of years, different countries in the Europe have tested different variants of Intelligent Speed Adaptation (ISA) systems to find the one with the best level of efficiency and acceptance [13]. A survey of literature shows many of these systems rely on the use of car sensors, smartphones, GPS and several image processing techniques [5]. Based on the level of permissiveness, ISA systems are classified as either informatory or intervening [14]. The informatory ISA only informs the driver of the speed limit permissible on the road but leaves the driver to decide whether the limit will be respected or not. The intervening ISA system mandatorily ensures the speed limits are respected either by cutting off fuel and air supply to the car engine or by making it impossible for the driver to maintain or increase pressure on the throttle once the speed limit is reached.

Many ISA systems rely heavily on GPS. For example, authors in [15] developed an intelligent speed adaptation system for company vehicles. The system utilizes a GPS receiver and a speed map to relay the speed limit on every road to the driver. A verbal warning is given to the driver if the speed limit is found to have been exceeded by more than 5 Km/hr. In cases of prolonged non-compliance, penalty points are allocated to the driver. The study found that using a reward system like this was able to reduce over speeding from 18.7% to 7.4 % in urban centers. GPS information was also used by authors in [16] to limit over speeding at bends. This problem arises from the fact that road geometry like curvatures are not usually factored in when road limits are set. The authors developed a dynamic speed adaptation system based on both road recommended speed limits as well as the road curvature information using the path tracking approach. Using GPS information, high curvature segments on roads as well as an appropriate speed limit for such curvature were harnessed to develop a robust database used to ensure safety for driver and passengers. GPRS-enabled GPS devices were also used in [17] to study the impact of auditory intelligent speed adaptation on the informal public transport system in south Africa. The study reported that by taking such approach, it was possible to reduce the occurrence of over speeding by more than 20 percentage points. As ubiquitous as GPS based systems are, it is important to note that error margins are usually non-trivial. In fact, error margin as large as 400m was reported in a GPS accuracy measurement study carried out in [18]. This large errors are connected with the line of sight requirement associated with GPS which to a large extent undermine the reliability of these systems particularly in tunnels and areas with high rise buildings.

Aside GPS, there are also several smart phone based designs reported in literatures. In [19], a smart phone based ISA system was developed to reduce vehicle fuel consumption on approach to traffic signal. To mitigate the common "hurry up and wait" attitude of drivers, the developed model by accurately predicting traffic signal movement can help to prevent drivers from making unnecessary acceleration. By using the traffic signal prediction model, the optimal speed of vehicles can be calculated which in turn translate into about 31.2 % savings on vehicles fuel consumption. A mobile application for road surface quality control model was proposed by [20]. The proposed model relies on the accelerometer in mobile phones coupled with GPS data to monitor and detect anomalies like pot holes and bumps on road surfaces with results showing huge prospect in using the application for automatic road quality sensing. Perera and Dias [14] also developed an intelligent mobile application that utilizes intelligent speed adaption system to help drivers avoid excessive speed in Sri Lanka. The application retrieves speed limit information from a speed limit database on a cloud server. The application then serves the driver audio and visual warning when limits are exceeded. For most of the smart phone based ISA systems, rapid draining of phone battery is a problem that makes prolong usage difficult [21].

Sensors and digital signal processing (DSP) based ISA systems have also been studied. Ultrasonic sensor was used to create an alerting system which warns a driver of the presence of a road bump in [11] with the loudness and frequency of the beep indicating the vehicles proximity to the bump. Infrared Sensor was used in [22]. In [23], light detection and ranging (LIDAR) was used to help autonomous vehicles detect obstacles on the road. The use of an intelligent sugeno fuzzy model to adapt the speed of a car to road curvature was proposed in [24]. As cheap as the usage of ultrasonic and infrared sensors are in ISA systems, camera based designs still remain one of the most common and efficient in practical applications. The model presented in [5] integrated a camera with DSP techniques to identify bumps on the road. The same approach was adopted in [25] to design real-time speed limit sign recognition system. Sensors and camera based approach are good but they do not have all the answers. Cameras can be affected by inclement weather conditions like fog, haze and poor illumination [23]. Given by the fact that most Nigerian highways do not have highway lighting, camera based systems may not suffice particularly when they are lacking in night vision capability. Ultrasonic sensors are also at best used as a support system since the possibility of the emitted sound wave being absorbed rather than reflected is likely [23]. Our specific contributions in this work are as follow:

(1) While there are several adaptive speed limiting framework in test, we specifically seek to address the associated risk of speed bumps on Nigerian highways. Knowing fully well that these bumps are to ensure vehicles travel at a near zero speed at these bump sites, we set out to use unidirectional radio waves communication to achieve this same objective. Instead of the 5.9 GHz band specified in the IEEE 802.11p standard, we used the 2.4 GHz ISM band due to spectrum restriction in Nigeria. The choice is premised on the finding by [26] which found that the results of using the DSRC band to be identical to that of the 2.4 GHz band;

(2) In our design, we showed that received signal strength at a receiver can be a suitable substitute to the use of GPS for ranging purpose;

(3) Since all vehicle in Nigeria are mandated by law to have speed limiter installed. More importantly, only specific models are approved for use by the FRSC. Our work was motivated by being able to integrate our design with what is already available.

## III. MODEL PARAMETERS

For traditional road bumps, the essence is to bring the vehicle to a near zero speed at bump location. The typical speed for vehicles at these bump locations usually range from about 4 km/hr to 8 km/hr. It is however important to note that there are other forms of road bumps that allow for much higher speed but these are not considered to present any significant threat or discomfort to either the vehicle or its occupants. For the developed model in this work, a speed of 6 km/hr which is the typical walking speed of most people was chosen as it is considered safe in most application. Depending on the payload data though, the maximum speed at the near zero zone can be more than 6 km/h but with a ceiling of 12 km/h which is considered the average jogging speed of most people.

Next, we model the stopping distance. In order to accommodate the various types of vehicles (sport cars, buses, trailers) that are expected on the roads as well as the possible road conditions (dry, wet or sloppy), extreme and worst possible scenarios typical in Nigeria have been assumed. The FRSC recommended speed ceiling of 120km/h was observed in the model development. This is the average upper boundary given that the poor condition of most roads in Nigeria do not permit a driver to maintain an average speed of more than 100 Km/hr. Another important model variable is the coefficient of friction. The variable depends largely on the nature of tread on the tyre as well as the nature of the road surface. For good quality tyre with good threads travelling on a dry road surface, coefficient of friction can be taken to be between 0.9 and 1. We assumed a little less than ideal scenario in which the tyre threads and road surface are average. The coefficient of friction in this circumstance was taken to be 0.7.

The average convenient stopping distance of a typical car on the highway can be calculated using (1). This is the distance a decelerating vehicle with initial velocity $u$ will cover before reaching a near-zero final velocity v.

$$s = \frac{u^2}{2\mu g} \quad (1)$$

In (1), s is stopping distance in meters, $\mu$ the coefficient of friction and g the convenient rate of deceleration which is taken to be $10 m/s^2$. The deceleration rate of $10 m/s^2$ is not arbitrarily but based on series of experiment performed on different vehicle types in [27]. The study revealed that a deceleration rate of $10 m/s^2$ is typical for most vehicles.

Using (1), the average stopping distance needed for a vehicle with initial velocity of 120 km/h, decelerating at a maximum rate of $10 m/s^2$ with 0.7 coefficient of friction is estimated to be 80 m. Consequently as shown in Fig. 1, the roadside unit (RSU) is placed 80m away from the near-zero speed location. Equation 2 shows the velocity-distance relationship of the vehicle. This models the average velocity of the vehicle as it decelerates from its initial velocity to the near zero final velocity v.

$$v = \sqrt{u^2 - 2g\mu s} \quad (2)$$

### A. Roadside Unit Model

The model for the Road Side Unit (RSU) is shown in Fig. 2. The model features an electrical power system (EPS), a communication system and an on-board computer (OBC) for data processing and control functions. The EPS comprises of solar panels and batteries for round the clock operation of the unit. The communication unit has both radio frequency (RF) and IoT gateway. A directional yagi antenna with a 20 degree beam width is proposed to ensure effective and reliable communication between the RSU and the IVU. The IoT gateway allows for remote monitoring and control of the transmission parameters over the internet in areas where such connectivity are available. This gives the option of disabling the electronic speed bump system whenever it is no longer needed. Other system parameters that can be varied over the IoT gateway include the desired speed at the bump site and the duration of time for which vehicles must travel at this speed before the limit is lifted. The FRSC which is the organization with oversight of road safety and security management matters in the country can monitor and adjust the system over an application server. In cases where adjustments are not initiated, the default settings of the system remains. For the RF transmission, the ISM band is proposed. Specifically for this work, the 2.4 GHz band is used. The on-board computer processes and control data flow primarily through the communication stack thus adding flexibility to the operation of the system.

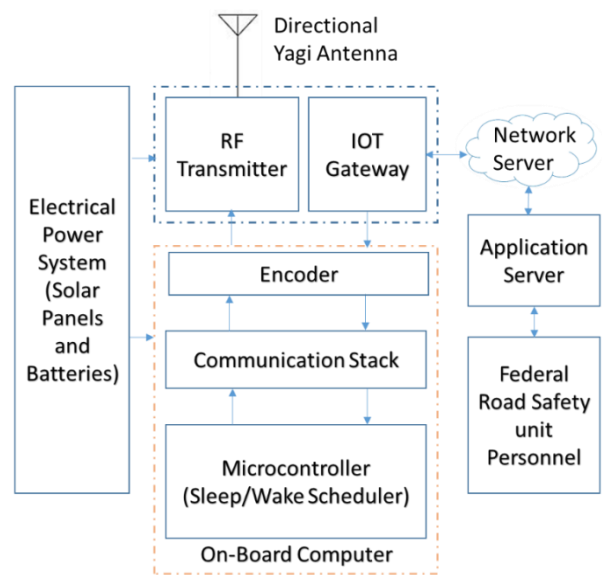

Figure 2. Electronic Speed Bump Road Side Unit

## B. Free Space Path Loss (FSPL)

The free space path loss, the link budget and the link margin was calculated using (3), (4) and (5) respectively.

$$FSPL = 20\log_{10}(d) + 20\log_{10}(f) + 20\log_{10}(\frac{4\pi}{c}) \quad (3)$$

$$P_{RX} = P_{TX} + G_{TX} - L_{TX} - L_{FS} - L_M + G_{RX} - L_{RX} \quad (4)$$

$$G_m = P_{RX} - R_s \quad (5)$$

Where d is the distance between the receiving and transmitting antenna, f the frequency, c the speed of light, $G_{TX}$ the gain of the transmitting antenna and $G_{RX}$ the gain of the receiving antenna, $L_{TX}$ the transmitter loss, $L_{RX}$ the receiver loss, $L_{FS}$ the free space loss, $G_m$ the link margin and $R_s$ the receiver sensitivity. By motivating the link design using (3) to (5), a minimum link range of 400m was ensured at all times in order to secure the model reliability.

## C. In-Vehicle Unit

The in-vehicle unit is the most important part of the system. The unit is located inside the vehicle and works together with the installed speed limiter to decelerate the vehicle from its initial speed to near-zero at the low speed zone. The unit and its operation protocol are shown in Fig. 3 and 4 respectively.

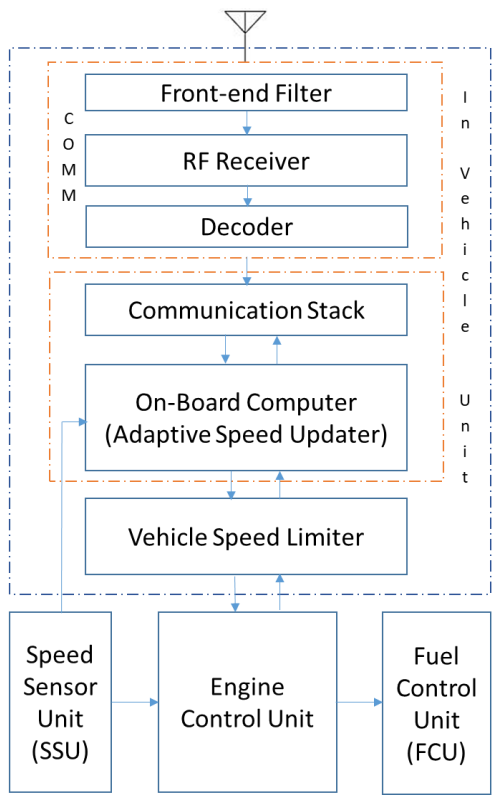

Figure 3. In-vehicle electronic speed breaker

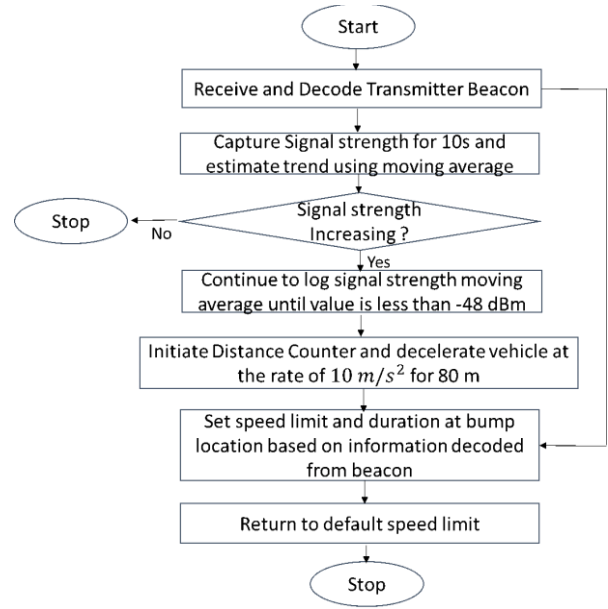

Figure 4. In-vehicle unit operation protocol

The rationale behind the design path is to ensure compatibility with existing speed limiting system mandated for all vehicles in the country. The IVU is made up of a bandpass filter, fixed frequency receiver and a decoder to decode the payload data. The receiver in the IVU is a fixed frequency receiver and hence is always locked to the RSU frequency. The IVU decoder decodes the payload data while the OBC uses the data to iteratively set the speed limit of the vehicle in order to slow it down.

From (3), the design estimated reception distance is a minimum of 400 m from the RSU. By calculating time as the ratio of distance to speed, the time lapse between the first reception of signal by vehicles travelling at different speed and reaching the RSU is estimated in Table 1.

From Table 1, the average duration between the time of intercepting transmission from the RSU and reaching it on the average is about 15 seconds which is enough to capture sufficient signal strength data as well as decoding payload data.

The IVU operation process is detailed in Fig. 4. Once the receiver intercepts the signal, it attempt decoding until successful and acquires the signal strength over 10 seconds. Since the channel is dynamic over the signal strength acquisition period, the signal trend is easily estimated using a 2$^{nd}$ order simple finite impulse response filter given by (6) to ensure multipath fading induced error is minimized.

TABLE I.   ESTIMATD TIME BETWEEN SIGNAL RECEPTION AND RSU

| S/N | Speed (km/h) | Time to reach RSU (secs) |
|---|---|---|
| 1 | 80 | 18.02 |
| 2 | 100 | 14.40 |
| 3 | 120 | 12.01 |

$$y[n] = \sum_{i=0}^{N} b_i\, x[n-i], \qquad (6)$$
$$b_i = \frac{1}{N+1}$$

Where x[n] are the received signal strength value at the receiver, y[n] the filtered output, N the filter order which is 2 and $b_i$ the impulse response at instant i.

The essence of the trend using the moving average finite impulse response filter is to help the IVU know whether it is moving away or approaching a RSU. If the IVU is on approach to a RSU, it uses the signal strength to determine its approximate position relative to the RSU. Using the signal strength rather than the GPS to calculate the distance between the RSU and IVU in order to establish a reference point from which deceleration distance count can be initiated is based on the observation that signal strength profile of radio frequency transmitter along roads is usually relatively stable over time [28]. Relying on this observation, we performed several simulations to find the average signal strength at the RSU which was found to be -40.03 dBm. To give margin for error, we set the reference point to -48 dBm. Consequently, once the receiver receives signal strength less or equal to -48 dBm, the receiver switch closes and the adaptive speed updater on the OBC iteratively adjust the maximum speed limit of the car at the rate of 10 m/s$^2$.

The RSU payload contains the speed limit as well as the distance over which this restriction must be complied with. In rare occasions, it is possible that the payload data were not successfully decoded, then the vehicle at the near zero limit zone will travel at the speed of 6 km/h over 20m after which the speed limit will be lifted and reset to the maximum allowable speed of 120 km/h obtainable on all roads in the country.

## IV. EXPERIMENTAL VALIDATION, RESULT AND DISCUSSION

The simulation parameters used for this work are presented in Table 2. In order to provide experimental validation for our work, we employ the dataset presented in [29].

TABLE II. MODEL SIMULATION PARAMETERS

| Simulation Parameter | Value |
|---|---|
| Transmitted Power | 10 dBm |
| Gain of Transmitting Antenna | 15 dBm |
| Losses from Transmitter | 5 dBi |
| Free Space Loss | -92.1 dB @ 400m |
| Miscellaneous Loss | 5 dB |
| Gain of Receiving Antenna | 8 dBi |
| Losses from Receiver | 5 dBm |
| Receiver's sensitivity | -90 dBm |
| Transmit Frequency | 2.4 GHz |

Though the focus of the study conducted in [29] is different from ours, we observed a similar approach with consistent experimental conditions including the ISM frequency band which sufficiently aligned with our work. More importantly, we also observed a similar signal strength trend which is typical of a vehicle approaching a road side RF transmitter. We equally correlated our results with these experimental data and observed correlation of 0.93 indicating a strong correlation. The result of both set of data is presented in Fig. 5. Observation from the simulation result reveals that the average signal strength around the location of the RSU is -40.03 dBm. This informed our choice of -48 dBm as the reference signal strength at which the receiver's switch closes. The maximum ranging error from this benchmarking was found to be 10 m.

An exponential increase in the received signal strength on the in-vehicle unit approach of the road side unit is observed in Fig. 5 indicating that this trend can be employed as a sufficient indication that the vehicle is moving towards the roadside unit.

In Fig. 6, the expected link margin of the IVU on approach to the RSU is shown.

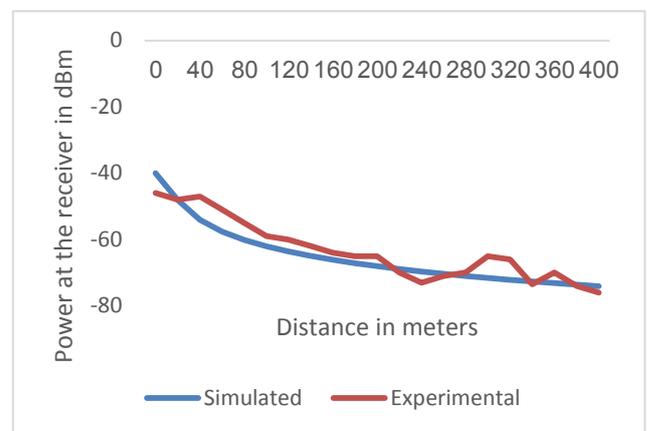

Figure 5. Received power at IVU agaist distance

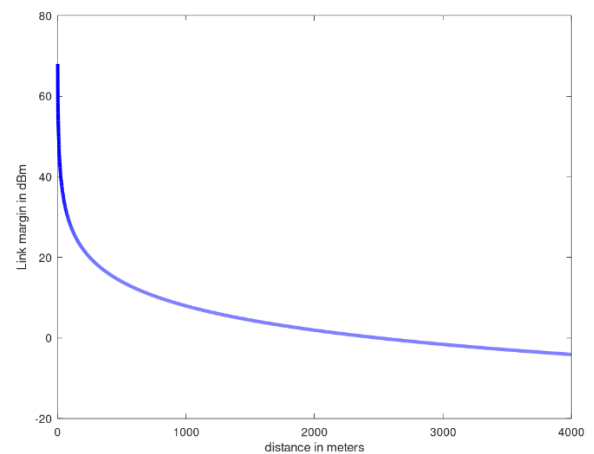

Figure 6. Link margin against distance

Theoretically, from the simulation result, the transmitter reach could be as far as four kilometers. This range will almost always be out of reach in practice but with a 16 dB link margin observed at a distance of 400m from the RSU, not only will there be ample resources for the IVU to ensure a successful communication and correct estimation of direction relative to the RSU but also to decode the RF payload shown in Table 3.

TABLE III. PAYLOAD DATA

| Header Files | Speed at Bump Site | Near zero speed duration |
|---|---|---|
| Header/CRC | 6 km/h | 20 m |

The vehicle deceleration profile is shown in Fig. 7 for vehicles travelling at the maximum allowable speed of 120 km/h. The vehicle can be seen to have decelerated to a near-zero speed just before the 80 km mark indicating the workability of the model.

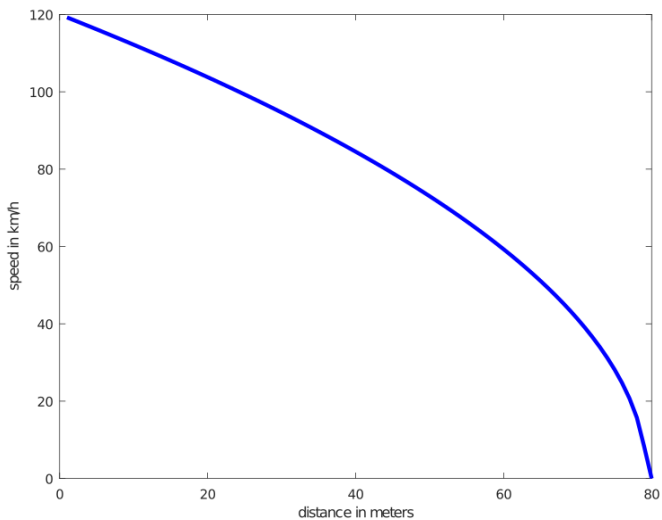

Figure 7. Deceleration profile of the vehicle over the stopping distance

## V. CONCLUSION

In this work, we specifically set out to develop a model that mimic the role of speed bumps on the road. Speed bumps are meant for residential and urban center roads where speed are not supposed to exceed 60 km/h. However in Nigeria, erecting bumps on highways are common sight. It has been established in several studies in Nigeria that speed bumps do not stop accidents, rather they cause them. In this work we further highlight some of the risks associated with speed bumps. In order to provide remedy to this hazard, we developed a radio frequency based speed bump. The motivation behind the design is to ensure that we take full advantage of the use of speed limiters that have been mandated for all vehicles in Nigeria. Our design consists of a road side unit and an in-vehicle unit. The RSU is located at a convenient 80m away from the intended speed bump site with the IVU located inside the vehicle. The RSU continuously sends beacons that contain information that is needed to regulate drivers speed at near zero speed zone. The receiver decodes the signal and uses the payload to adaptively reduce the speed of the vehicle until a near zero speed is attained at the bump site. In our design, we propose a model that adaptively set the speed limit of a speed limiter to eradicate the need for road bumps using the broadcast mode of communication. Finally, we proofed that received signal strength at the receiver can be a reliable substitute to GPS based adaptive speed limiting models.

ACKNOWLEDGMENT

Special thanks to the Federal University of Technology Akure for providing all the support needed for the successful completion of this research.